\documentclass[a4paper]{jpconf}

\usepackage{float}
 
\usepackage{graphicx}

\usepackage{amsmath, amsthm, amssymb, amsfonts}

\usepackage{lineno}

\usepackage{subfigure}
\usepackage{bm}% bold math
\usepackage[percent]{overpic}
\usepackage{color}

\usepackage{hyperref}
\usepackage[all]{hypcap}
\hypersetup{
    colorlinks,
    citecolor=blue,
    filecolor=blue,
    linkcolor=black,
    urlcolor=blue
}

\def\pt{p_{\rm T}}

\def 	\C2{C_{\rm 2}}

\def\sNN{\mbox{$\sqrt{s_{_{\rm NN}}}$}}

\usepackage[pscoord]{eso-pic}% The zero point of the coordinate systemis the lower left corner of the page (the default).

\newcommand{\placetextbox}[3]{% \placetextbox{<horizontal pos>}{<vertical pos>}{<stuff>}
  \setbox0=\hbox{#3}% Put <stuff> in a box
  \AddToShipoutPictureFG*{% Add <stuff> to current page foreground
    \put(\LenToUnit{#1\paperwidth},\LenToUnit{#2\paperheight}){\vtop{{\null}\makebox[0pt][c]{#3}}}%
  }%
}%
\begin{document}
\title{Angular correlations with charmed hadrons in the Monte-Carlo model with string repulsion}

\author{I G Altsybeev$^1$, G A Feofilov$^1$}
\address{$^1$ 
%Saint-Petersburg State University, 7/9 Universitetskaya Emb., 199034, Saint Petersburg, Russia
Saint-Petersburg State University, 7/9 Universitetskaya nab., St. Petersburg, 199034 Russia
}

\ead{Igor.Altsybeev@cern.ch}

\begin{abstract}
Recent experimental results revealed large elliptic flow of the charmed hadrons at LHC energies, unexpectedly  similar to that of the charged pions. The mentioned measurements are often interpreted using transport models, which incorporate dissociation and recombination mechanisms for charm quarks. In this  report, a modified version of the Monte Carlo model with string repulsion is used to calculate %rapidity and 
azimuthal correlations with charmed hadrons. The string repulsion mechanism may provide significant angular correlations and can be considered as an alternative to thermalization picture.
\end{abstract}

% #######################
\section{Introduction}

\placetextbox{0.88}{0.075}{1}

This is usually argued that
heavy quarks (charm and beauty) are predominantly produced at the very early stages of a heavy-ion collision via the hard scattering processes  and, therefore, they experience all subsequent stages of the system evolution (see some recent reviews in \cite{1,2}). 
 %For example, in QGP ... rescatterings...
%One evidence for that %of the interactions of heavy quarks with the %medium 
%comes from the observation of 

A  rather strong modification of the transverse momentum ($\pt$) distributions of heavy-flavour hadrons in heavy-ion collisions in comparison to pp collisions, was observed in the experiments at RHIC \cite{PHENIX, STAR} and LHC (see \cite{3} for ALICE data).  It may be considered as one of the indications of the interactions of heavy quarks with the medium. 
A large suppression of heavy-flavour hadron yields for $\pt > 4$-5 GeV/$c$
 in central  collisions is usually attributed to 
parton energy loss in the medium
% [SOME REF].
\cite{E-loss-1, E-loss-2, E-loss-3}.
Another evidence comes from the recent measurements of
the second Fourier coefficient $v_2$ (called an elliptic flow)
of D mesons in Pb--Pb collisions at LHC energies ($\sNN$=2.76 and 5.02 TeV) \cite{D_v2_ALICE_502}.
It was found that the magnitude of the $v_2$ turned out to be similar to that of charged pions.
These measurements are usually compared with  theoretical model calculations which implement charm-quark transport in a hydrodynamically expanding medium.
% and have the potential to constrain medium parameters.

% ###########################################
\section{The MC model with string repulsion}

In the current work, we 
%will 
try to qualitatively describe the behaviour of D mesons
in the framework of the Monte Carlo (MC) model with string repulsion.
%Namely, 
We will try to separate so-called  ``soft" and ``hard" sources of particle production, 
and will see how the  interplay between these two contributions
is reflected on some observables. 
% will allow 
%where the soft contribution comes from string breaking mechanism. % (...).
%The idea is ... of this soft and hard components

The ``soft" part of  particle production comes from the hadronization of the quark-gluon strings.
The strings are formed at  the early stages of hadron-hadron collisions, they may overlap in case of sufficiently high densities and interact by repelling or attracting each other, depending on the direction of the color fluxes \cite{abramovsky}. For instance, in the hypothesis of repulsive interaction, strings may acquire, before the hadronization, an additional transverse boost. This produces additional transverse momenta to the particles formed in string decays over a wide range of rapidity, thus leading to modification of observables.
Details of the Monte Carlo implementation of this model are described in \cite{IA_string_repulsion_MC},
and results on azimuthal asymmetry of two-particle correlations
obtained with this model
are shown in  \cite{IA_GF_Quark2016}.

In order to parametrize  the  spectra of particles
coming from breaking of a string in its rest frame,
the identified particle spectra of pions, kaons and protons measured by ALICE \cite{production_pi_K_p_pp_PbPb_276}
in proton-proton collisions at $\sqrt{s}=2.76$ TeV were used.
Namely, for each particle species, the exponential fit for the soft part of the spectrum  (usually at $\pt<1$ GeV/$c$)
was performed, and then utilized in the MC model to assign the transverse momentum for the particles emitted from the strings.
%while t
The remaining part of each spectrum, obtained after subtraction of the exponential part,
was used as a ``hard" component for corresponding particle species.
This component comes from the hard scatterings of partons,
and for high $\pt$  % ($> 10$ GeV/$c$) 
the spectra have a power-law shape, as expected from the perturbative QCD % (pQCD)
calculations. 

The important feature of the model presented here is that the hard part 
{\it is not affected} by the transverse boosts of the strings, acquired due to the repulsive interaction.
As a consequence, in the non-peripheral (i.e. central) nucleus-nucleus collisions, where strings accumulate significant 
transverse boosts, the soft component of the spectrum
is being ``blue-shifted" onto the hard one, leading to  the modification of the overall shapes of the spectra in nucleus-nucleus collisions, simulated in this MC model.

There is a very small probability for the quark-gluon string to break into 
charm and anti-charm quark pair via the Schwinger mechanism ($\sim 10^{-11}$, \cite{Gen_purpuse_MC_generators}).
Nevertheless, we assume in the present MC model, that %this probability is not negligible,
$c$-$\overline{c}$ quark pairs can be produced in the string
hadronization process with a sufficient probability,
providing the exponential part of the spectra of charmed hadrons, which is supplemented by the hard part. Such growth of the open charm production can be explained by the modification of the string tension due to string overlapping and interaction in a form of fusion. String breaking into $c$-$\overline{c}$  quark pairs  was studied in  \cite{vk_SQM}.

As a proxy for the charmed hadrons,
D mesons were taken, and their spectrum measured in pp collisions by ALICE \cite{D_spectra_RAA} was used to parametrize the MC model, in the same manner
as spectra of $\pi$, K, p.
% the string with much higher probability...

For the simplified version of the MC model used in this work, it is supposed also that it is possible to neglect  resonance decays, jets and  final state effects like rescattering of hadrons.

% #########################################################
\section{Nuclear modification factor for A--A collisions}

\placetextbox{0.88}{0.075}{2}

% ######## Fig  RAA
\begin{figure}[!t]
\centering
\begin{overpic}[width=0.58\textwidth]{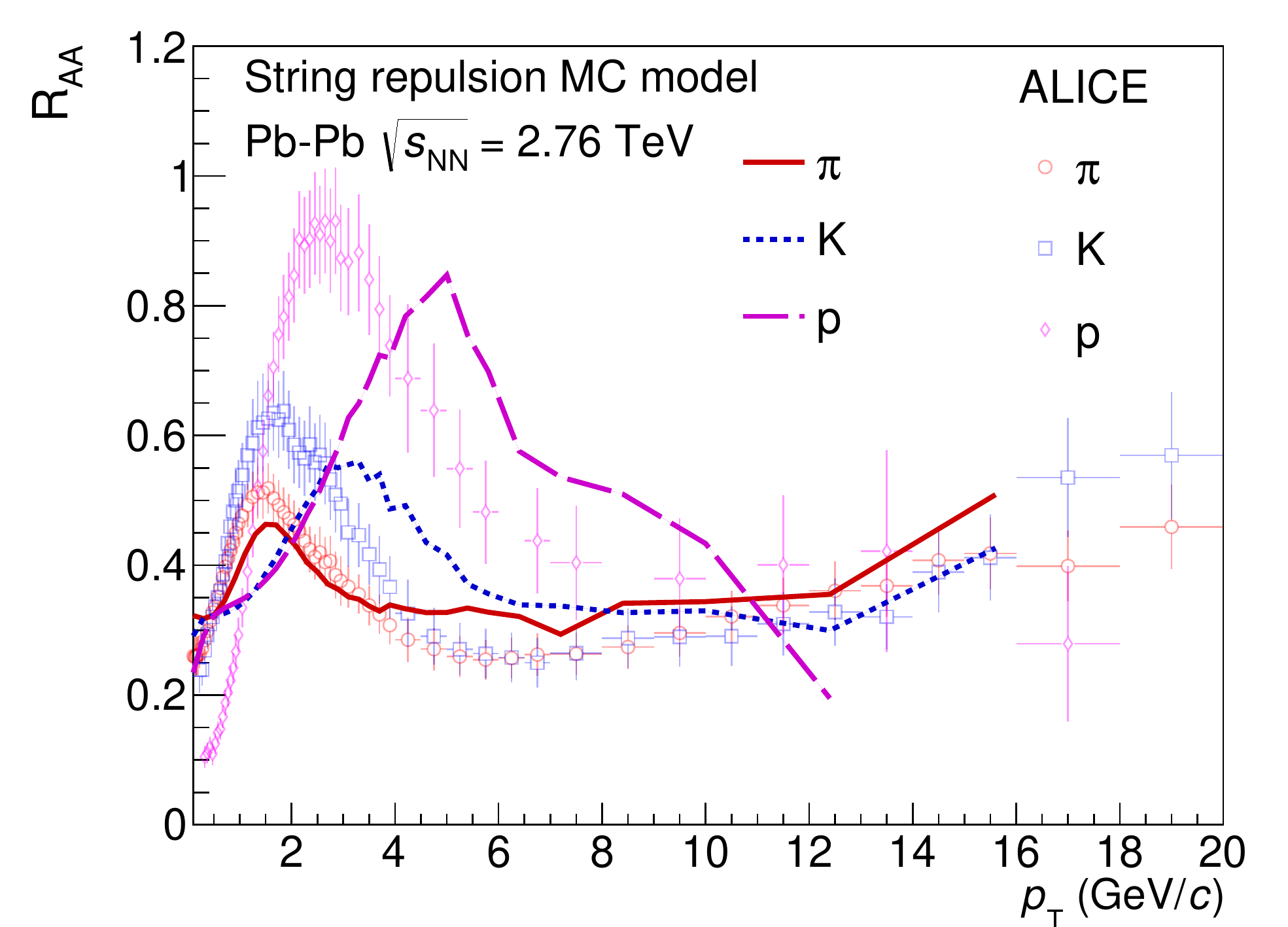}
          \put(17.5,55){  centrality 20--40\%}
\end{overpic}
\caption{
$R_{\rm AA}$ for Pb--Pb collisions in centrality range of 20--40\%.
Model calculations (lines) are compared with ALICE data (markers).
}
\label{fig:RAA} 
\end{figure}

In order to compare to the experiment,
it is interesting to capture the $\pt$--dependence of the nuclear modification  factor ($R_{\rm AA}$) in the framework of the  string repulsion model.
The $R_{\rm AA}$ factors, calculated within the MC model for Pb--Pb collisions in centrality range of 20--40\%
for pions, kaons and protons, are shown in Figure \ref{fig:RAA} as lines of different colors.
Markers represent values measured by ALICE~\cite{RAA_pi_K_p}.
Correspondence between the model and data is only qualitative.
However, it can be seen that the so-called mass ordering is correctly reproduced by the MC model,
and also
 a decreasing trend is observed after some maximum value
 for each particle species, similarly to data.
This change of behavior reflects the interplay between the soft and hard components,
implemented in the MC model.

Some differences for pions at small $\pt$  could  come from decays of  resonances, which are switched off in the model.
Peaks for $\pi$, K and p 
in the model calculations are shifted to the right with respect to data points,
indicating that radial expansion  is slightly stronger in the simulation
than in data.
Also, the known increasing trend of $R_{\rm AA}$,
observed in data  for $\pt>8$~GeV/$c$  up to very high $\pt$ values,
is not  captured in the given simulations, since we do not introduce any modification of the ``hard" component in this variant of the model.

% #######################
\section{Differential elliptic flow}

The mass ordering of particle species was observed also
in the differential elliptic flow, $v_2$($\pt$), for instance
%was observed
 in Pb-Pb collisions \cite{v2_pi_K_p_ALICE}.
Recently, the $v_2$($\pt$) has been measured also for D meson 
%was also measured 
in semi-central Pb--Pb events~\cite{D_v2_ALICE_502},
and the values were found to be similar to that of lighter particles.

% ######## Fig  v2
\begin{figure}[!b]
\centering
\begin{overpic}[width=0.58\textwidth]{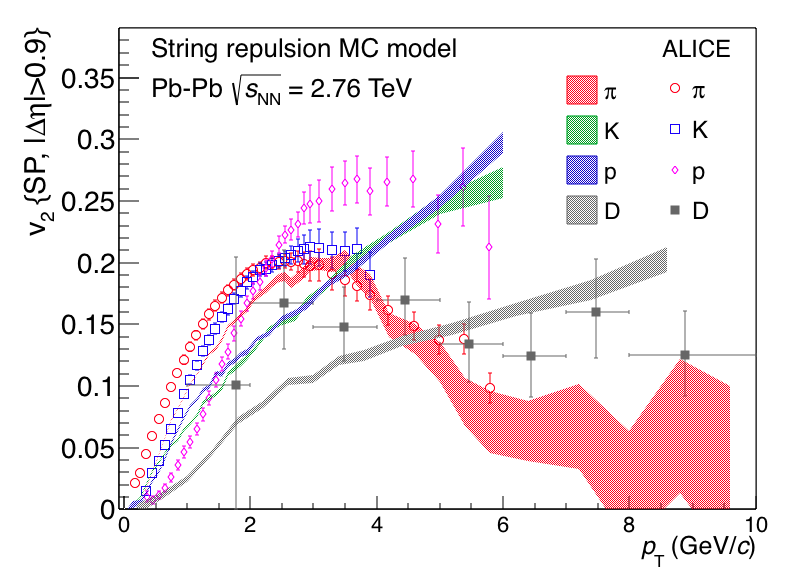}
          \put(17.5,55){  centrality 20--40\%}
\end{overpic}
\caption{
Differential elliptic flow for $\pi$, K, p and D mesons
in centrality range of 20--40\% in the MC model.
Markers are for  ALICE data (for D mesons, centrality range in experiment
was 30--50\%).
}
\label{fig:v2} 
\end{figure}

\placetextbox{0.88}{0.075}{3}

In the present MC model, the mass ordering of $v_2$ obtained for different particle species is qualitatively similar to the experiment (Figure \ref{fig:v2}). 
%Change of the trend to lower values at high $\pt$
%As in was seen for  for $R_{\rm AA}$, shown in Figure~\ref{fig:RAA}, 
% one may see that 
Data points for D mesons are also fairly described in the model.
For pions, after some maximum value,
one may see a decreasing  trend of $v_2$  at high $\pt$,
as it was observed for $R_{\rm AA}$ in Figure \ref{fig:RAA}.
This is  due to  increasing  fraction of the hard contribution
 at higher $\pt$, which is not affected
by the ``collective" repulsive motion of the strings.
In order to see the same change of behavior 
for other particle species, more subtle tuning of the interplay between
soft and hard components of the particle production is needed.
%to decrease towards smaller $v_2$ values at higher $\pt$. 

% #######################
\section{Conclusions}

It is shown that some certain signatures of collectivity, like $R_{\rm AA}$  or elliptic flow,  observed experimentally in A--A collisions, can be obtained in the model with the repulsing strings. Namely, the model qualitatively reproduces the %experimentally observed 
mass ordering of the nuclear modification factor and the azimuthal flow both  for the light flavours and for heavier hadrons, for instance, D mesons. 

%(assuming possibility for a string to hadronize into charmed hadrons).

The main challenge for the model 
%to get better quantitative agreement with data 
is to reproduce correctly the  interplay between soft and hard contributions in the intermediate $\pt$ range (from $\approx 1$ to 8 GeV/$c$, depending on the particle species).

\placetextbox{0.88}{0.075}{4}

%\vspace*{-0.5cm}
\section*{Acknowledgements}
This work is supported by the Russian Science Foundation, 
grant 16-12-10176.

% #######################

\end{document}